\documentclass[aps,prl,twocolumn,showpacs,groupedaddress,amsmath,amssymb]{revtex4}

\usepackage{graphicx}

\begin{document}

\bibliographystyle{apsrev}

\title{A model for codon position bias in RNA editing}

\author{Tsunglin Liu and Ralf Bundschuh}

\affiliation{Department of Physics, Ohio State University, 191 W Woodruff Av., Columbus OH 43210-1117}

\begin{abstract}
RNA editing can be crucial for the expression of genetic information via 
inserting, deleting, or substituting a few nucleotides at specific 
positions in an RNA sequence. Within coding regions in an RNA sequence, 
editing usually occurs with a certain bias in choosing the positions of 
the editing sites. In the mitochondrial genes of {\it Physarum polycephalum}, 
many more editing events have been observed at the third codon position 
than at the first and second, 
while in some plant mitochondria the second codon position dominates. 
Here we propose an evolutionary model that explains this bias as the basis 
of selection at the protein level. The model predicts a distribution of 
the three positions 
rather close to the experimental observation in {\it Physarum}. 
This suggests that the codon position bias in {\it Physarum} is mainly 
a consequence of selection at the protein level.  
\end{abstract}

\pacs{87.23.Kg, 05.40.-a, 87.14.Gg, 87.14.Ee}

\maketitle

\section{Introduction}

The central dogma of molecular biology states that the transfer of
genetic information flows from the genomic DNA to messenger RNA to
proteins. This implies that there are two fundamental processes
involved in the fabrication of proteins. The first such process is
transcription and it consists of copying the genomic DNA into a
messenger RNA of identical sequence. Both, the genomic DNA and the
messenger RNA use an alphabet of four letters which makes direct
copying of one sequence into the other possible. The second process is
called translation and consists of synthesizing a protein, i.e., a
sequence of amino acids, following the instructions contained in the
sequence of the messenger RNA. During translation groups of three
consecutive bases of the messenger RNA --- the codons --- are read in
order to determine which of the $20$ amino acids is to be appended to
the protein being synthesized. The map from the $4^3=64$ possible
codons into the $20$ possible amino acids is called the genetic code.

Before translation into proteins, it has been discovered that RNAs
transcribed from most eukaryotic genes undergo a variety of processing
events, that convert RNA precursors into mature RNAs ready for
translation.  For example, the splicing process removes extended
stretches of the nucleotide sequences called introns from an RNA
precursor such that only the remaining RNA sequence codes for a
protein.

Besides these normal processing events, many novel phenomena that
change the content of RNA sequences before translation have been
discovered in several different organisms~\cite{grosj98,bass00,benne86,simps89,mahen91,gott93,gualb89,covel89,hiesel89,powell87,chen87}. 
These phenomena, which are
now coined as RNA editing, consist of inserting, deleting, or changing
individual or a very small number of nucleotides. Still, even by
changing only a few nucleotides RNA editing can significantly alter
the coding and result in functionally distinct proteins. In addition,
RNA editing has also been found to occur in non-coding regions like tRNAs
and rRNAs, and can alter the function of these RNA molecules as well.

For editing within the coding regions, there is often a significant
codon position bias. For example, it has been discovered that in the
mitochondrial genes of the slime mold {\it Physarum polycephalum},
about two thirds of the editing events that insert Cs happen at the
third positions of their respective codons~\cite{mahen91,gott93}. On the 
contrary, in mitochondrial genes of {\it Arabidopsis} and other plants, 
about 90 percent of the editing events, which in this case convert Cs to 
Us, happen at the first two positions of the codons~\cite{giege99,handa03,notsu02}.

Codon position bias is somewhat surprising since there is no obvious 
relation
between the editing which happens on the RNA level and codons which in
principle have a meaning only during translation. The underlying reason
why the editing machinery might prefer certain codon positions for editing
is still not understood. 

Here, we propose an evolutionary model which explains the 
editing position bias in mRNAs. In the evolutionary scheme, a
population of an organism undergoes mutations which alter the 
DNA sequences of the individuals in the population
and change their genetic information.
Under natural selection, the members in the gene pool that carry the genes 
with higher fitness grow faster and increase their relative frequency
in the total population.
 
The fitness of a gene in general depends on the encoded protein sequence
as well as other biological parameters.
In our model we will assume that the only dominant selection mechanism 
is the fitness of the resulting protein sequence. This 
assumption is reasonable if editing is relatively inexpensive, because in this 
case most editing sites will be random in nature rather than involve some 
other biological factors. Comparing our results with the actual codon 
position bias 
data then reveals if this assumption is true or if there are other, more 
fundamental selection mechanisms at work in a given organism. 

The two cases we apply our model to are the mitochondrial mRNAs of 
{\it Physarum polycephalum} and of the plants  {\it Arabidopsis thaliana}, 
{\it Brassica napus}, and {\it Oryza sativa}. Abundant editing events 
have been observed in these organisms. In {\it Physarum} mitochondrial 
mRNAs, one in every 25 bases is edited on average, 
which leads to about 1 in every 8 codons being edited on average. In 
plant mitochondrial mRNAs, about 2\% of the nucleotides are edited on 
average. 
In the remainder of this article, we briefly describe the editing events in 
these organisms. Then we focus first on the evolutionary model for 
{\it Physarum}. Following similar approaches for the case of {\it Physarum}, 
we then move on to the case of plant mitochondria.

Mitochondrial RNAs of {\it Physarum polycephalum} have been found to be 
edited extensively by insertions of mono-(C,U) and dinucleotides 
(CU,GU,UA,AA,GC, and UU)~\cite{mahen91,gott93}. Among the editing events 
within mRNAs of {\it Physarum polycephalum}, C insertions are
the most frequent events. In plant mitochondria, the most abundant 
events are C to U conversions. Thus, we only focus on these most frequent 
editing events in this communication.

For C insertions in 11 mitochondrial mRNAs of 
{\it Physarum polycephalum}~\cite{parimi}, 
some editing positions are ambiguous as nucleotide Cs are inserted right 
next to another C. Excluding these ambiguous editing sites, 227 C 
insertions are observed within the coding regions. Among these, 58 and 24 
insertions occur at the first and second positions of the codons, respectively.
The remaining 145 insertions are found at the third codon position. In this case, 
the third position is the most favorable and the second position is the 
least favorable as already noted in previous work comprising only 5 
mitochondrial mRNAs~\cite{miller93}. 

In plant mitochondria, the dominant editing event is a 
C to U conversion. Thus, there is no ambiguity in determining the codon 
position of these events. The editing codon position preferences of the 
three plant mitochondria studied here are quite similar. In {\it Arabidopsis}, 154/236/51 
C to U conversions have been observed at the three positions~\cite{giege99}. 
In {\it Oryza}~\cite{handa03} and {\it Brassica}~\cite{notsu02}, 
142/243/33 and 174/230/77 conversions 
have been observed respectively. This suggests that the editing position bias 
stems from the same mechanism for the three organisms. Notice that the 
order of the position bias is the opposite to that of {\it Physarum}. 

We now develop our model for the insertional editing events of 
{\it Physarum}. This model will later be easily adapted to 
the case of substitutional editing such as in the plant mitochondria.

The scheme of our evolutionary model is the following. During the 
proliferation of {\it Physarum}, nucleotide mutations occur at random 
positions 
in the mitochondrial DNA sequence. These mutations can be substitutions, 
insertions, and deletions. Although we model all three types of mutations, 
we are mostly interested in deletions in the DNA sequence. Most offspring 
with a deletion die because the protein produced according to the mutated 
DNA sequence is out of frame from the site of the deletion on and thus 
cannot function properly. However, some few deletions may be accepted by 
the editing machinery which would insert back nucleotides C to those 
positions of mutation and thus rescue them. Under natural selection, 
these offspring would survive if the resulting protein could function 
properly in place of the original one.

The genetic code, i.e., the rules of translation from codons to amino 
acids, is organized such that often the third codon position is irrelevant 
to the identity of the amino acid interpreted into the protein sequence. 
Thus, the third codon position is the least sensitive to nucleotide changes 
to C generated by editing events. Therefore, in this mutation-selection 
model, random deletions at the third codon position survive much better 
than at the first and second position.

\begin{figure}
\includegraphics[width=8cm]{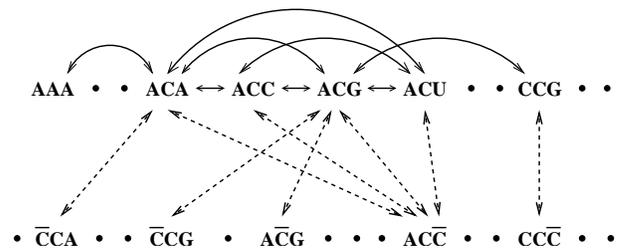}
\caption{Mutations among codons of the evolutionary model for {\it Physarum}. 
Solid lines stand for mutations between regular codons. Dashed 
lines are for mutations between a regular codon and a codon with an 
editing site.}
\label{mutate}
\end{figure}

Beyond this qualitative picture, the random mutations and selection can be 
rigorously formulated in the following way using as an example the codon ACG. 
For random mutations, we assume the substitutions and deletions occur 
randomly at one of the three positions with certain mutation rates $\mu_s$ 
and $\mu_e$. The random substitutions result in 9 regular codons by 
converting the base in each position into 3 other nucleotides 
(Fig.~\ref{mutate}). The random 
deletions generate 3 codons with editing sites $\bar{C}$CG, or A$\bar{C}$G, 
or AC$\bar{C}$, where the bar stands for a vacant site in the DNA sequence 
after mutation. As an amino acid is expressed, this vacant site will be 
treated as a C to represent the editing event in which a nucleotide C is 
inserted before translation. Note that $\mu_e$ is the effective rate 
for a deletion that is accepted by the editing machinery and rescued 
by inserting a C. This rate is in general smaller than the base 
deletion rate. For codons with a vacant site, they also 
undergo insertions which randomly insert one of the four nucleotides 
(A,C,G,U) back into the vacant site with a mutation rate $\mu_e$, which
we take to be the 
same as the random deletion rate. 

The basic assumption of our model is that selection happens only at the 
amino acid level. Thus, the fitness of a codon, i.e., the growth 
rate, depends only on the similarity of the new amino acid coded for by 
the mutated codon to the original amino acid. To be specific, we will use the 
amino acid scoring matrix BLOSUM62~\cite{henik92} to quantify the 
similarities between 
amino acids. However, our results are not very sensitive to the choice of 
this matrix. As an example, the initial codon ACG is assigned a 
large growth rate of 6 because it translates into the desired amino acid 
threonine. The mutated codon $\bar{C}$CG would express the amino acid proline and is 
assigned a growth rate of -1 since proline behaves quite differently from thrieonine 
according to the BLOSUM62 matrix.  

Apart from the fitness at the amino acids level, a codon created by an 
editing event is considered as less fit than a regular one because the 
editing machinery presumably uses up resources in order to perform the
editing. We thus reduce the growth rate for a codon created by an editing 
event by an editing cost $c$.

In order to cast this model in mathematical terms
we define the fraction of codon $i$ as $x_i$, where $i=1\sim 64$ is for the 
64 regular codons and $i=65\sim 112$ is for all codons containing an 
editing site. The corresponding growth rates and editing costs are defined as 
$g_i$ and $c_i$, where $g_i$ is obtained from the BLOSUM62 matrix and $c_i=c$ 
for $i=65\sim 112$ and zero otherwise. We then write the general model as 
\begin{eqnarray}
\dot{x_i}\!&=&\!\!\left(g_i\!-\!c_i\!-\!\sum_{j\ne i}\mu_{ij}\right)\!x_i\!+\!\sum_{j\ne i} 
\mu_{ji}x_j\!-\!x_i\!\left(\sum_j g_j x_j\right)\! \nonumber \\
&=& \sum_j M_{ij}x_j-x_i\left(\sum_j g_j x_j\right)
\end{eqnarray} 
where $\mu_{ij}$ is the general mutation rate from codon $i$ to 
codon $j$, and is set as $\mu_s$ for substitutions, $\mu_e$ for deletions 
and insertions, and $0$ for codons that cannot be mutated into each other.

We will look at the distribution of regular codons and codons with 
editing sites in this model at equilibrium. Although the last term in 
this equation is non-linear, Eigen's theory~\cite{josh04,thomp73} 
tells us that the equilibrium distribution is proportional to the 
eigenvector of the matrix 
$M_{ij}$ for the largest eigenvalue. Thus, obtaining the equilibrium 
distribution is merely a matter of numerically finding the eigenvector 
for the largest eigenvalue of a $112\times 112$ matrix which we do with 
Mathematica~\cite{math}. 

To obtain some insight into this evolutionary model, we again first look at 
the case where the initial codon is ACG. We will focus 
on the limit of strong selection in which most results become largely 
independent of the individual mutation rates $\mu_s$ and $\mu_e$. We 
obtain this strong selection limit by setting $\mu_s$ and $\mu_e$ 
to be much smaller than 1 since the growth rates that set the time scale 
are of order 1. To be more specific, we set $\mu_s=\mu_e=10^{-6}$ because 
the relative magnitude of $\mu_s$ to $\mu_e$ only slightly affects the 
result. 
As expected, 
the equilibrium distribution shows that only those codons that translate 
into the original amino acid survive. Thus, among the codons with editing 
sites, only A$\bar{C}$A, A$\bar{C}$C, A$\bar{C}$G, A$\bar{C}$U, 
AC$\bar{C}$ 
survive since the genetic code depends only on the first two positions in 
this case. We also find that the ratio between editing at 
the second position and at the third position is about one. Thus, we arrive at 
a simple scheme in the strong selection limit that only the mutations that 
result in the same amino acid survive and each position that survives 
contributes equally at equilibrium. 

This simple scheme allows us to quickly predict the ratio among edited 
positions for each codon. The overall ratio is then easily obtained by 
considering all possible codons with proper weights. The weights of the 
codons are determined by the experimentally observed unedited codon 
frequencies in the DNA 
sequence. This scheme results in the ratio of editing positions as 
17/11/72 percent which is rather close to the experimental observed 
ratio 25/11/64 percent.

The above result is obtained without any editing cost, i.e., $c$=0. We 
find that the role of the editing cost is mainly to reduce the 
fraction of the total number of 
codons with editing sites; the ratio among the editing positions is only 
slightly changed upon introducing an editing cost. Thus, by adjusting the 
editing cost, we can tune the fraction of codons with C insertions 
to match the experimental result of about 7\%. In this case, 
the overall ratio of editing positions is about 19/15/66 percent, 
which is still very close to the experimental result.

We conclude that this simple evolutionary model provides a possible 
scenario for the editing codon position bias. This suggests that in 
{\it Physarum} the majority of editing events might indeed be subject to no 
other biological constraint but the fitness of the resulting protein 
sequence. Thus, editing events might be randomly acquired. 

This is also consistent with the evolution of RNA editing in {\it Physarum} 
and its close relatives~\cite{horton00}. Abundant editing has also 
been observed in the 
organisms {\it Didymium nigripes} and {\it Stemonitis flavogenita}. 
However, in {\it Arcyria cinerea} and {\it Clastoderma debaryanum}, 
only few C insertions or even none are observed. This indicates that 
the editing machinery in {\it Arcyria} and {\it Clastoderma} is not yet
fully developed. The fact that {\it Physarum} has acquired so many editing 
sites since the divergence from {\it Arcyria} and {\it Clastoderma} suggests 
that editing in {\it Physarum} is less expensive and that thus random editing 
is likely. 

To study plant mitochondrial genes, we slightly modify our
evolutionary model as follows. The random substitutions between regular 
codons remain the same as in the case of {\it Physarum}, and they occur with a 
rate $\mu_s$. As to the editing events that convert Cs to Us, we assume 
that for certain bases that mutate to Cs, the editing machinery would 
recognize these sites and convert them to Us before translation. Here, 
we use $\bar{U}$ to represent such a mutation. Thus, as an example, the 
codon ACG can mutate into $\bar{U}$CG, A$\bar{U}$G, and AC$\bar{U}$ as 
the corresponding nucleotide mutate to a C and then is recognized by the 
editing machinery. This process involving 
an editing event occurs with a mutation rate $\mu_e$. Again, the back 
mutation that converts a $\bar{U}$ to one of the four nucleotides (A,C,G,U) 
is modeled to also occur with the same mutation rate $\mu_e$. Mathematically, the 
evolutionary model for {\it Physarum} and plant mitochondria are identical. 

This evolutionary model, when applied to plant mitochondrial genes, does 
not predict the correct preference in the codon positions for editing. 
Instead, it still predicts the majority of editing sites to occur at the third 
codon position.  
This suggests that in plant mitochondrial genes editing events do not
just happen randomly, but that there exist
some other biological mechanisms that bias the choice of editing sites. 

This assertion is consistent with what is known about the editing mechanism 
in subcellular organelles of plants. Recent experiments show that 
different editing sites in subcellular organelles of plants 
usually require different proteins in the editing process~\cite{miya02}. In 
such a case, each editing site is quite expensive in terms of resources 
the cell has to provide. It is highly unlikely that such expensive 
editing sites are randomly acquired. The existence of editing 
events in spite of the fact that they are expensive
is an indication that these editing sites are significantly involved 
in biological functions beyond simply providing the correct protein 
sequence. 


In summary, the closeness of the result of our evolutionary model to the 
experimental observations in {\it Physarum} suggests that the editing 
position bias in {\it Physarum} is mainly a consequence of random mutations with selection 
at the protein level. In the case of plant mitochondria, the disagreement 
between the evolutionary model and the observations implies that some 
other biological factors besides selection at the protein level must 
also play a role in the evolution, so that the random mutation-selection 
scheme does not fit in these organisms.

We gratefully acknowledge contributions by Taeyoung Choi
to the very early stages of this project and fruitful discussions
with Jonatha Gott. This work has
been supported by grant no. DMR-0404615 from the National
Science Foundation.


\begin{thebibliography}{99}

\bibitem{grosj98}
H. Grosjean, R. Benne, eds. 1998. Modification and Editing of RNA, 
Washington, DC: ASM Press.

\bibitem{bass00}
B. Bass, ed. 2000. RNA editing: Frontiers in Molecular Biology. 
Oxford Univ. Press

\bibitem{benne86}
R. Benne {\it et al.}, Cell {\bf 46}, 819 (1986)

\bibitem{simps89}
L. Simpson and J. Shaw, Cell {\bf 57}, 355 (1989) 

\bibitem{mahen91}
R. Mahendran, M. R. Spottswood and D. L. Miller, Nature {\bf 349}, 434 (1991)

\bibitem{gott93}
J. M. Gott, L. M. Visomirski, and J. L. Hunter, J. Biol. Chem. {\bf 268}, 
25483 (1993)

\bibitem{gualb89}
J. M. Gualberto {\it et al.}, Nature {\bf 341}, 660 (1989)

\bibitem{covel89}
P. S. Covello and M. W. Gray, Nature {\bf 341}, 662 (1989)

\bibitem{hiesel89}
R. Hiesel, B. Wissinger, W. Schuster, and A. Brennicke, Science {\bf 246}, 
1632 (1989)

\bibitem{powell87}
L. M. Powell {\it et al.}, Cell {\bf 50}, 831 (1987)

\bibitem{chen87}
S. H. Chen {\it et al.}, Science {\bf 238}, 363 (1987)

\bibitem{giege99}
P. Gieg\'e and A. Brennicke, PNAS {\bf 96}, 15324 (1999)

\bibitem{handa03}
H. Handa, Nucl. Acids. Res., {\bf 31}, 20 (2003)

\bibitem{notsu02}
Y. Notsu {\it et al.}, Mol. Genet. Genomics {\bf 268}, 434 (2002)

\bibitem{parimi}
N. Parimi, R. Bundschuh and J. M. Gott (to be published)

\bibitem{miller93}
D. Miller {\it et al.}, Sem. Cell Biol. {\bf 4}, 261 (1993)

\bibitem{henik92}
S. Henikoff and J. G. Henikoff, PNAS.{\bf 89}, 10915 (1992)

\bibitem{josh04}
J. B. Plotkin, J. Dushoff, M. M. Desai, H. B. Fraser,  q-bio.PE/0410013 (2004)

\bibitem{thomp73}
C. J. Thompson and J. L. McBride, Math. Biosci. {\bf 21}, 127 (1973)

\bibitem{math}
Wolfram Research, Inc., Mathematica, Version 5.0, Champaign, IL (2003).

\bibitem{horton00}
T. L. Horton and L .F. Landweber, RNA {\bf 6}, 1339 (2000)

\bibitem{miya02}
T. Miyamoto, J. Obokata, M. Sugiura, Mol. Cell Biol. {\bf 22}, 6726 (2002)

\end{thebibliography}
\end{document}